\documentclass[epj]{svjour}
\usepackage{graphics}
\usepackage{epsfig}

\def\lQ{\Lambda_{\rm QCD}}

\newcommand{\be}{\begin{equation}}
\newcommand{\ee}{\end{equation}}
\newcommand{\bea}{\begin{eqnarray}}
\newcommand{\eea}{\end{eqnarray}}

\def\als{\alpha_{\rm s}}
\def\siml{{\ \lower-1.2pt\vbox{\hbox{\rlap{$<$}\lower6pt\vbox{\hbox{$\sim$}}}}\ }}
\def\simg{{\ \lower-1.2pt\vbox{\hbox{\rlap{$>$}\lower6pt\vbox{\hbox{$\sim$}}}}\ }}
\newcommand{\MS}{\overline{\rm MS}}

\begin{document}
\title{Heavy Quarkonium Physics from Effective Field Theories}
\author{Antonio Vairo 
\thanks{antonio.vairo@mi.infn.it}}
\institute{Dipartimento di Fisica dell'Universit\`a di Milano and INFN, 
via Celoria 16, 20133 Milano, Italy}
\date{Received: 19 October 2006}
\abstract{
I review recent progress in heavy quarkonium physics from an effective 
field theory perspective. In this unifying framework, I discuss advances in perturbative 
calculations for low-lying quarkonium observables and in lattice calculations for 
high-lying ones, and progress and lasting puzzles in quarkonium production.
}

\maketitle
\section{Introduction}
\label{intro}
\begin{figure}
\makebox[0cm]{\phantom b}
\put(0,0){\epsfxsize=8truecm\epsffile{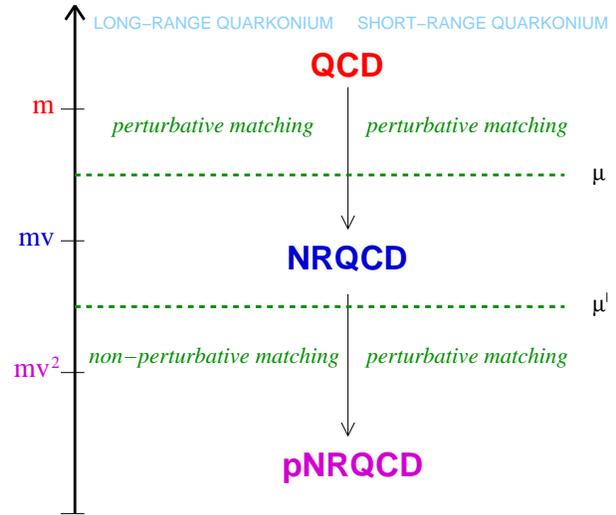}}
\caption{EFTs for quarkonium.}
\label{fig:scales}       
\end{figure}

Heavy quarkonium is most frequently and successfully stu\-died 
as a non-relati\-vi\-stic bound state of QCD \cite{Brambilla:2004wf}.
Non-relativistic bound sta\-tes made of heavy quarks are characterized by a hierarchy of energy scales:
$m$, $m v$, $m v^2$, ..., $m$ being the heavy quark mass and $v$ the heavy quark 
velocity in the centre-of-mass frame. A way to disentangle rigorously these scales is
by substituting QCD, scale by scale, with simpler but equivalent Effective Field Theories (EFTs).
Modes of energy and momentum of order $m$ may be integrated out from QCD 
in a perturbative manner ($m \gg \lQ$, by definition of heavy quark) leading 
to an EFT known as non relativistic QCD (NRQCD) \cite{Caswell:1985ui,Bodwin:1994jh}; integrating out gluons of energy or 
momentum of order $mv$ leads to an EFT known as potential NRQCD (pNRQCD) \cite{Pineda:1997bj}, 
see Fig. \ref{fig:scales}. pNRQCD is close to a Schr\"odinger-like description of the bound
system and, therefore, as simple. The bulk of the interaction
is carried by potential-like terms, but non-potential interactions,
associated with the propagation of low-energy dynamical degrees of freedom, are generally present as well.
For a review on the subject we refer to \cite{Brambilla:2004jw}. EFTs for quarkonium were discussed at the 
conference by J. Soto.

QCD is characterized by an intrinsic scale $\lQ$. Let us consider states below threshold.
In the perspective of non-relativistic EFTs, it may be important to distinguish between low-lying quarkonium 
resonances, for which we assume $mv^2 \simg \lQ$, and  high-lying states, for which we assume
$mv \sim \lQ$. In the first case, the system is weakly coupled and the potential is perturbative, 
in the second case, the system is strongly coupled and the potential must be determined 
non-perturbative\-ly, for instance, on the lattice. This difference is not appreciated 
in potential models that typically describe the whole spectrum with the same interaction.
Besides potential terms, which encode the dynamics at the soft scale $mv$, EFTs contain also 
terms that describe the dynamics of lower energy  degrees of freedom.  These terms are typically 
absent in potential models. For what concerns systems close or above the open flavor threshold, 
a complete and satisfactory understanding of their dynamics has not yet been achieved. 
Therefore, the study of these systems, despite their phenomenological interest, is on a less 
secure ground than the study of states below threshold. Indeed, it largely relies 
on phenomenological models. 

In the following, I will outline some recent (and very recent) progress in our theoretical understanding
of heavy quarkonium phy\-sics. The perspective is quite personal, not only because 
the framework, which I will privilege, is that one of QCD EFTs, but also because
I can only present a selection of results.
It is important, when discussing about progress, to keep in mind the distinction 
made above about low-lying, high-lying and threshold sta\-tes. In the first case, 
we may rely on perturbation theory: the challenge is in performing higher-order 
calculations and the goal is precision physics. In the second case, we have to rely 
on non-perturbative methods: the challenge is in providing a consistent framework 
where to perform lattice calculations and progress is measured by the progress in lattice 
computations. In the third case, one of the major challenges is to interpret the new charmonium
states discovered at the B-factories in the last few years
(for an updated list, see http://www.qwg.to.infn.it/).
The distinction may help to better appreciate the quality of the progress made.

\section{Low-lying $Q\bar{Q}$}
Low-lying $Q\bar{Q}$ states are assumed to realize the hierarchy: $ m \gg  mv
\gg mv^2 \simg \lQ$, where $mv$ is the typical scale of the inverse distance between the
heavy quark and antiquark and $mv^2$ the typical scale of the binding energy.
At a scale $\mu$ such that $mv \gg \mu \gg mv^2$ the effective degrees of freedom are 
$Q\bar Q$ states (in color singlet and octet configurations), low-energy gluons and light quarks. 

The lowest-lying quarkonium states are $\eta_b$ (not yet detected), $\Upsilon(1S)$, $\eta_c$ and  
$J/\psi$. The $\Upsilon(1S)$ and $J/\psi$ masses may be used to determine 
the bottom and charm quark masses. These determinations 
are competitive with other ones (for the $b$ mass see e.g. \cite{El-Khadra:2002wp}).
We report some recent determinations in Tab. 1.

\begin{table}[h]
\addtolength{\arraycolsep}{0.2cm}
\begin{center}
\begin{tabular}{|c|c|c|}
\hline
reference & order &  ${\overline m}_b({\overline m}_b)$ (GeV) 
\\
\hline
\cite{Pineda:2001zq} & NNNLO$^*$ & $4.210 \pm 0.090 \pm 0.025$
\\
\cite{Brambilla:2001qk} & NNLO +charm & $4.190 \pm 0.020 \pm 0.025$
\\
\cite{Eidemuller:2002wk} & NNLO& $4.24 \pm 0.10$
\\
\cite{Penin:2002zv} & NNNLO$^*$& $4.346 \pm 0.070$
\\
\cite{Lee:2003hh} & NNNLO$^*$ & $4.20 \pm 0.04$
\\
\cite{Contreras:2003zb} & NNNLO$^*$ & $4.241 \pm 0.070$ 
\\
\cite{Pineda:2006gx} & NNLL$^*$ & $4.19 \pm 0.06$ 
\\
\hline
\hline
reference  & order &  ${\overline m}_c({\overline m}_c)$ (GeV)  
\\
\hline
\cite{Brambilla:2001fw} & NNLO & $1.24 \pm 0.020$ 
\\
\cite{Eidemuller:2002wk}  & NNLO & $1.19 \pm 0.11$ 
\\
\hline
\end{tabular}
\vspace{2mm}
\caption{Different recent determinations of ${\overline m}_b({\overline m}_b)$
and ${\overline m}_c({\overline m}_c)$ in the $\MS$ scheme from the bottomonium and the 
charmonium systems. The displayed results either use direct determinations or non-relativistic 
sum rules. Here and in the text, the $^*$ indicates that the theoretical input is only partially 
complete at that order.}
\end{center}  
\label{Tabmasses}
\end{table} 

Once the heavy quark masses are known, one may use them to extract other
quarkonium ground-state observables. At NNLO the $B_c$ mass was calculated 
in \cite{Brambilla:2000db} ($M_{B_c} = 6326 \pm 29$  MeV), 
\cite{Brambilla:2001fw} ($M_{B_c} = 6324 \pm 22$  MeV) and 
\cite{Brambilla:2001qk} ($M_{B_c} = 6307 \pm 17$  MeV). These values 
agree well with the unquenched lattice determination of \cite{Allison:2004be} 
($M_{B_c} = 6304 \pm 12^{+18}_{-0}$ MeV), which shows that the $B_c$ mass
is not very sensitive to non-perturbative effects. 
This is confirmed by a recent measurement of the $B_c$ in the channel  $B_c \to J/\psi \, \pi$ 
by the CDF collaboration at the Tevatron; they obtain with 360 pb$^{-1}$ of data 
$M_{B_c} = 6285.7 \pm 5.3 \pm 1.2$ MeV \cite{Acosta:2005us}, while the latest
available figure based on 1.1 fb$^{-1}$ of data is 
$M_{B_c} = 6276.5 \pm 4.0 \pm 2.7$ MeV 
(see http://www-cdf.fnal.gov/physics/new/bottom/060525.ble \\ ssed-bc-mass/).

The bottomonium (and charmonium) ground-state hyperfine splitting has been calculated at NLL
in \cite{Kniehl:2003ap}. Combining it with the measured $\Upsilon(1S)$ mass,
this determination provides a quite precise prediction for the $\eta_b$ mass: 
$M_{\eta_b} = 9421 \pm 10^{+9}_{-8} ~{\rm MeV}$, where the first error is an estimate of the
theoretical uncertainty and the second one reflects the uncertainty in $\als$.
Note that the discovery of the $\eta_b$ may provide a very competitive 
source of $\als$ at the bottom mass scale with a projected error at the $M_Z$ 
scale of about $0.003$. Similarly, in \cite{Penin:2004xi}, 
the hyperfine splitting of the $B_c$ was calculated at NLL
accuracy: $M_{B_c^*}  - M_{B_c} = 65 \pm 24^{+19}_{-16}~{\rm MeV}$.

The ratio of electromagnetic decay widths was calculated for the ground state 
of charmonium and bottomonium at NNLL order in \cite{Penin:2004ay}.
In particular, they report: $\Gamma(\eta_b\to\gamma\gamma) /\Gamma(\Upsilon(1S)\to e^+e^-) = 0.502
\pm 0.068 \pm 0.014$, which is a very stable result with respect to scale variation.
A partial  NNLL$^*$ order analysis of the absolute width of $\Upsilon(1S) \to
e^+e^-$ can be found in \cite{Pineda:2006ri}.

Allowed magnetic dipole transitions between charmonium and bottomonium ground states
have been considered at NNLO  in \cite{Brambilla:2005zw,Vairo:2006js}.
The results are: $\Gamma(J/\psi \to \gamma \, \eta_c) \! = (1.5 \pm 1.0)~\hbox{keV}$
and $\Gamma(\Upsilon(1S) \to \gamma\,\eta_b)$ $=$  $(k_\gamma/39$ $\hbox{MeV})^3$
$\,(2.50 \pm 0.25)$ $\hbox{eV}$, where  the errors account for uncertainties (which
are large in the charmonium case) coming from higher-order corrections.
The width $\Gamma(J/\psi \to \gamma\,\eta_c)$ is consistent with 
\cite{Yao:2006px}. Concerning $\Gamma(\Upsilon(1S) \to \gamma\,\eta_b)$, 
a photon energy $k_\gamma = 39$ MeV corresponds to a $\eta_b$ mass of 9421 MeV. 

The radiative transition $\Upsilon(1S)\to\gamma\,X$ has been considered in 
\cite{Fleming:2002sr,GarciaiTormo:2005ch}. The agreement with the CLEO data of 
\cite{Nemati:1996xy} is very satisfactory. J. Soto has reported about this at the conference.

\section{Low-lying $QQq$}
The SELEX collaboration at Fermilab reported evidence of five resonances that 
may possibly be identified with doubly charmed baryons  \cite{Ocherashvili:2004hi}. 
Although these findings have not been confirmed by other experiments (notably 
by FOCUS, BELLE and BABAR) they have triggered a renewed theoretical interest in doubly heavy baryon systems.

Low-lying $QQq$ states are assumed to realize the hierarchy: $ m \gg mv \gg
\lQ$, where $mv$ is the typical inverse distance between the two heavy quarks
and $\lQ$ is the typical inverse distance between the centre-of-mass of the two heavy quarks
and the light quark. At a scale $\mu$ such that $mv \gg \mu \gg \lQ$ the effective  
degrees of freedom are $QQ$ states (in color antitriplet and sextet
configurations), low-energy gluons and light quarks. The most suitable EFT at
that scale is a combination of pNRQCD and HQET
\cite{Brambilla:2005yk,Fleming:2005pd}. The hyperfine splittings of the doubly heavy 
baryon lowest states have been calculated at NLO in $\als$ and at LO in
$\lQ/m$ by relating them to the hyperfine splittings of the $D$ and $B$ mesons (this 
method was first proposed in \cite{Savage:di}). In \cite{Brambilla:2005yk}, the 
obtained values are: $M_{\Xi^*_{cc}}-M_{\Xi_{cc}} = 120 \pm 40$ MeV 
and $M_{\Xi^*_{bb}}-M_{\Xi_{bb}} = 34 \pm 4$ MeV, which are 
consistent with the quenched lattice determinations of 
\cite{Flynn:2003vz,Lewis:2001iz,AliKhan:1999yb,Mathur:2002ce}.
Chiral corrections to the doubly  heavy baryon masses, strong decay widths and 
electromagnetic decay widths have been considered in \cite{Hu:2005gf}.

Also low-lying $QQQ$ baryons can be studied in a weak coupling framework.
Three quark states can combine in four color configurations: a singlet,
two octets and a decuplet, which lead to a rather rich dynamics
\cite{Brambilla:2005yk}. Masses of various $QQQ$ ground states have been 
calculated with a variational method in \cite{Jia:2006gw}: since baryons made of three 
heavy quarks have not been discovered so far, it may be important for future searches  
to remark that the baryon masses turn our to be lower 
than those generally obtained in strong coupling analyses.

\section{High-lying $Q\bar{Q}$}
High-lying $Q\bar{Q}$ states are assumed to realize the hierarchy: $ m \gg  mv \sim \lQ
\gg mv^2$. A first question is where the transition from low-lying to high-lying takes place.
This is not obvious, because we cannot measure directly $mv$. Therefore, the 
answer can only be indirect and, so far, there is no clear agreement in the literature.
A weak-coupling treatment for the lowest-lying 
bottomonium states ($n=1$, $n=2$ and also for the $\Upsilon(3S)$) appears 
to give positive results for the masses  at NNLO in \cite{Brambilla:2001fw}
and at N$^3$LO$^*$ in \cite{Penin:2005eu}.
The result is more ambiguous for the fine splittings of the 
bottomonium $1P$ levels in the NLO analysis of \cite{Brambilla:2004wu} and positive only 
for the $\Upsilon(1S)$ state in the  N$^3$LO$^*$ analysis of \cite{Beneke:2005hg}. 
In the weak-coupling regime, the magnetic-dipole hindered transition $\Upsilon(2S) \to \gamma\,\eta_b$ 
at leading order \cite{Brambilla:2005zw} does not agree with the experimental upper bound 
\cite{Artuso:2004fp}, while the ratios for different $n$ of the radiative decay widths 
$\Gamma(\Upsilon(nS) \to \gamma\,X)$ are better consistent with
the data if $\Upsilon(1S)$ is assumed to be a weakly-coupled bound state 
and $\Upsilon(2S)$ and $\Upsilon(3S)$ strongly coupled ones \cite{GarciaiTormo:2005bs}. 

Masses of high-lying quarkonia may be accessed by lattice calculations.
A recent unquenched QCD determination of the charmonium spectrum below the open flavor threshold 
with staggered sea quarks may be found in \cite{Gottlieb:2005me}. At present, bottomonium is too heavy to be implemented 
directly on the lattice. A solution is provided by NRQCD \cite{Lepage:1992tx}. 
Since the heavy-quark mass scale has been integrated out, 
for NRQCD on the lattice, it is sufficient to have a lattice spacing $a$ as coarse as $m \gg 1/a \gg mv$.
A price to pay is that, by construction, the continuum limit cannot be reached. 
Another price to pay is that the NRQCD Lagrangian has to be supplemented by matching coefficients 
calculated in lattice perturbation theory, which encode the contributions from the 
heavy-mass energy modes that have been integrated out. A recent unquenched determination 
of the bottomonium spectrum with staggered sea quarks can be found in \cite{Gray:2005ur}.
Note that all matching coefficients of NRQCD on the lattice are taken at their tree-level value. 
This induces a systematic effect of order $\als v^2$ for the radial splittings 
and of order $\als$ for the fine and hyperfine splittings. In \cite{Gray:2005ur}, also 
the ratio $\Gamma(\Upsilon(2S)\to e^+e^-)/\Gamma(\Upsilon(1S)\to e^+e^-)\times M_{\Upsilon(2S)}^2/M_{\Upsilon(1S)}^2$ 
has been calculated. The result on the finest lattice compares well with the 
experimental one.

In order to describe electromagnetic and hadronic inclusive decay widths 
of heavy quarkonia, many NRQCD matrix elements are needed. 
The specific number depends on the order in $v$ of the non-relativistic 
expansion to which the calculation is performed and on the power counting.
At order $mv^5$ and within a conservative power counting, 
$S$- and $P$-wave electromagnetic and hadronic decay widths for bottomonia and charmonia below threshold 
depend on 46 matrix elements \cite{Brambilla:2002nu}.
More are needed at order $mv^7$ \cite{Bodwin:2002hg,Ma:2002ev,Brambilla:2006ph}.
Order $mv^7$ corrections are particularly relevant for $P$-wave quarkonium 
decays, since they are numerically as large as NLO corrections 
in $\als$, which are known since long time \cite{Barbieri:1980yp} and to which the most recent 
data are sensitive \cite{Vairo:2004sr,Brambilla:2004wf}.
NRQCD matrix elements may be fitted to the experimental decay data
\cite{Mangano:1995pu,Maltoni:2000km} or calculated on the lattice
\cite{Bodwin:1996tg,Bodwin:2001mk,Bodwin:2005gg}. The matrix elements of
color-singlet operators can be related at leading order to the
Schr\"odinger wave functions at the origin \cite{Bodwin:1994jh} 
and, hence, may be evaluated by means of potential models
\cite{Eichten:1995ch} or potentials calculated on the lattice
\cite{Bali:2000gf}.  However, most of the matrix elements remain 
poorly known or unknown. We refer to \cite{Brambilla:2004wf} for a summary of results.

At a scale $\mu$ such that $mv \sim \lQ \gg \mu \gg mv^2$, confinement 
sets in. Far from threshold, the effective degrees of freedom are
$Q\bar Q$ states (in color singlet configuration) and light quarks. 
Neglecting light quarks, the $Q\bar Q$ propagation is simply described 
by a non-relativistic potential \cite{Brambilla:2000gk,Pineda:2000sz}.
This will be in general a complex valued function admixture of perturbative 
terms, inherited from NRQCD, which encode high-energy contributions, 
and non-perturbative ones. The latter may be expressed in terms of Wilson loops 
and, therefore, are well suited for lattice calculations. 

The real part of the potential has been one of the first quantities to be calculated 
on the lattice (for a review see \cite{Bali:2000gf}).
In the last year, there has been some remarkable progress.
In \cite{Koma:2006si}, the $1/m$ potential has been calculated for the first time.
The existence of this potential was first pointed 
out in the pNRQCD framework \cite{Brambilla:2000gk}. 
A $1/m$  potential is typically missing in potential 
model calculations. The lattice result shows that the potential has a  
$1/r$ behaviour, which, in the charmonium case, 
is of the same size as the $1/r$ Coulomb tail of the static potential 
and, in the bottomonium one, is about 25\%. 
Therefore, if the $1/m$ potential has to be considered part 
of the leading-order quarkonium potential together with the static one, 
as the pNRQCD power counting suggests and the lattice seems to show, 
then  the leading-order quarkonium potential would be, somewhat surprisingly, a flavor-dependent function.
In \cite{Koma:2006fw}, spin-dependent potentials have been calculated 
with unprecedented precision. In the long range, they show, for the first 
time, deviations from the flux-tube picture of chromoelectric 
confinement \cite{Buchmuller:1981fr} (for a review see, for instance, \cite{Brambilla:1999ja}).
The knowledge of the potentials in pNRQCD could provide an alternative to the
direct determination of the spectrum in NRQCD lattice simulations:
the quarkonium masses would be determined by solving the Schr\"odinger
equation with the lattice potentials. The approach may present some advantages: 
the leading-order pNRQCD Lagrangian, differently from the NRQCD one, 
is renormalizable, the potentials are determined once for ever for
all quarkonia, and the solution of the Schr\"odinger equation provides also the 
quarkonium wave functions, which enter in many quarkonium observables: ~
decay widths,~ transitions,~ production cross-sections, ... .

The imaginary part of the potential provides the NR\-QCD decay matrix elements
in pNRQCD. They typically factorize in a part, which is the wave function in
the origin square (or its derivatives), and in a part which contains 
gluon tensor-field correlators \cite{Brambilla:2001xy,Brambilla:2002nu,Brambilla:2003mu,Vairo:2003gh}.
This drastically reduces the number of non-perturbative parameters needed; in pNRQCD,
these are wave functions at the origin and universal gluon tensor-field correlators, 
which can be calculated on the lattice.
Another approach may consist in determining the correlators on one set 
of data (e.g. in the charmonium sector) and use them to make predictions 
for another (e.g. in the bottomonium sector). Following this line in 
\cite{Brambilla:2001xy,Vairo:2002nh}, at NLO in $\als$, but at leading 
order in the velocity expansion, it was predicted 
${\Gamma_{\rm had}(\chi_{b0}(2P))}/{\Gamma_{\rm had}(\chi_{b2}(2P))} \approx$
 $4.0$ and ${\Gamma_{\rm had}(\chi_{b1}(2P))}/$ ${\Gamma_{\rm
had}(\chi_{b2}(2P))} \approx$ $0.50$. Both determinations turned 
out to be consistent, within large errors, with the CLEO III data 
\cite{Brambilla:2004wf}.

\section{Threshold states}
For states near or above threshold a general systematic treatment
does not exist so far. Also lattice calculations are inadequate. Most of the
existing analyses rely on models (e.g. the Cornell coupled channel model \cite{Eichten:1978tg}
or the $^3P_0$ model \cite{LeYaouanc:1972ae}).

However, in some cases, one may develop an EFT owing to special dynamical
conditions. An example is the $X(3872)$ discovered by BELLE \cite{Choi:2003ue} 
and seen also by CDF \cite{Acosta:2003zx}, D0 \cite{Abazov:2004kp} and BABAR \cite{Aubert:2004ns}. 
If interpreted as a loosely bound $D^0 \, \bar{D}^{*\,0}$ and  
${\bar D}^0 \, D^{*\,0}$ molecule, one may take advantage of
the hierarchy of scales $\lQ \gg m_\pi$ $\gg m_\pi^2/(2 m_{\rm red})$ 
$ \! \approx \! 10 ~{\rm MeV}$ $\gg E_{\rm binding}$. Indeed, the binding energy, 
$E_{\rm binding}$, which may be estimated from $M_{X(3872)} - (M_{D^{0\,*}}+M_{D^{0}})$,
is very close to zero, i.e. much smaller than the natural scale 
$m_\pi^2/(2 m_{\rm red})$. Systems with a short-range interaction 
and a large scattering length have universal properties that may be exploited; 
in particular, production and decay amplitudes factorize in a short-range 
and a long-range part, where the latter only depends on one single parameter, 
the scattering length \cite{Braaten:2003he,Braaten:2005jj,AlFiky:2005jd}.

Another interesting case is provided by the $Y(4260)$, discovered by BABAR  
\cite{Aubert:2005rm} and seen also by CLEO \cite{Coan:2006rv} and BELLE \cite{BELLE:2006fj}.
If interpreted as a heavy charmonium hybrid (see e.g. \cite{Kou:2005gt}), 
one may rely on the heavy-quark expansion and on lattice calculations 
to study its properties. In particular, analogously to the energy of a static quark-antiquark 
pair, the energies of static hybrids have been calculated in quenched approximation 
on the lattice in \cite{Juge:2002br}.
One may expect that the static energy is related to the static potential 
of the system and that this may be a relevant quantity for the dynamics of the system,  
but to substantiate this a suitable EFT for heavy hybrids, like pNRQCD for heavy quarkonium, 
needs to be formulated. Such a formulation does not exist yet.\footnote{
It is suggestive, however, that the lowest hybrid state 
is expected to contain a pseudoscalar color-octet quark-antiquark
pair and gluons, whose quantum numbers are those of the electric cloud in a 
diatomic $\Pi_u$ molecule, so that the system has $J^{PC}$ numbers $1^{--}$, like the $Y$.
Moreover its mass, obtained by solving the Schr\"odinger equation, is consistent, within 
100 MeV, with the experimental range 4.25-4.30 GeV.}

\section{Production}
Although a formal proof of the NRQCD factorization formula for heavy quarkonium production 
has not yet been developed, NRQCD factorization has proved to be very successful 
to explain a large variety of quarkonium production processes (for a review see 
the production chapter in \cite{Brambilla:2004wf}).
In the last year, there has been a noteworthy progress toward an all order proof.
In \cite{Nayak:2005rt,Nayak:2006fm}, it has been shown that a necessary condition 
for factorization to hold at NNLO is that the conventional octet NRQCD production matrix elements 
must be redefined by incorporating Wilson lines that make them manifestly gauge 
invariant. 

Differently from decay processes, a pNRQCD treatment does not exist so far for 
quarkonium production. The difficulty of such a formulation may be linked 
to that of providing a full proof of factorization at the level of NRQCD and 
a consistent definition of the NRQCD production matrix elements at the level of pNRQCD. 
A pNRQCD formulation of quarkonium production may present potentially the same 
advantages as that of quarkonium decay: a sensible reduction in the number 
of parameters and hence more predictive power. 

In the last years, two main problems have plagued our understanding 
of heavy quarkonium production:
(1) double charmonium production in $e^+e^-$ collisions and 
(2) charmonium polarization at the Tevatron.

In \cite{Abe:2004ww}, BELLE measures 
$\sigma(e^+e^-\to J/\psi+\eta_c) \, {\rm Br}(c\bar{c}\to$ $> \hbox{2 charged}) 
= 25.6\pm 2.8\pm 3.4~\hbox{fb}$ and in \cite{Aubert:2005tj}, BABAR finds 
$\sigma(e^+e^-\to J/\psi+\eta_c) \, {\rm Br}(c\bar{c} \to$ $> \hbox{2 charged}) 
= 17.6\pm 2.8^{+1.5}_{-2.1}~\hbox{fb}$. When these cross sections first appeared, 
they were about one order of magnitude above theoretical expectations. 
In the meantime, some errors have been corrected in some of the theoretical 
determinations, and, more important, NLO corrections in $\als$ have been calculated in 
\cite{Zhang:2005ch} and higher-order $v^2$ corrections in \cite{Bodwin:2006dn}.
All these improvements have shifted the theoretical value much closer to 
the experimental one. In \cite{Lee:QWG2006}, a preliminary 
estimate of $\sigma(e^+e^-\to J/\psi+\eta_c)$ that includes the above corrections 
has been presented, it reads: $ 16.7\pm 4.2~\hbox{fb}$.
It appears that, within the uncertainties, the discrepancy has been resolved. 
Still open is the issue of the inclusive double charm production in the presence
of a $J/\psi$. BELLE measures a ratio 
$\sigma(e^+e^-\to J/\psi+ c\bar{c})/\sigma(e^+e^-\to J/\psi+ X)$ that is about 
80\%, to be compared with theoretical estimates, which are about 10\% 
(see \cite{Brambilla:2004wf} for a detailed discussion). New theoretical analyses 
are timely.

Charmonium polarization has been measured at the Tevatron by the CDF collaboration 
at run I on 110 pb$^{-1}$ of data \cite{Affolder:2000nn} and recently at run II 
on 800 pb$^{-1}$ of data \cite{Kim:QWG2006}.
The data of the two runs do not seem consistent with each other 
in the 7-12 GeV region of transverse momentum, $p_T$, and both 
are not with NRQCD expectations. For large $p_T$, NRQCD predicts that 
the main mechanism of charmonium production is via color-octet gluon fragmentation, 
the gluon is transversely polarized and most of the gluon polarization 
is expected to be transferred to the charmonium.
For an analysis of the NRQCD prediction and its dependence 
on the adopted power counting, we refer to \cite{Fleming:2000ib}.
The CDF data do not show any sign of transverse polarization at large $p_T$.
Before drawing definite conclusions, a polarization
study from the D0 experiment would be most welcome, at least to settle 
the possible discrepancy between the run I and run II data.

\section{Conclusions}
Many new data on heavy-quark bound states are provided in these years by the 
B-factories, CLEO, BES, HERA and the Tevatron experiments. Many more will come 
in the future from the LHC and GSI. They will show new (perhaps exotic) states, 
new production and decay mechanisms. What ma\-kes all this interesting 
is that we may investigate a wide range of heavy quarkonium observables 
in a controlled and systematic fashion and, therefore, learn about one of the most elusive sectors 
of the Standard Model: low-energy QCD.
The tools for this systematic investigation are  provided by EFTs and lattice gauge 
theories. Still challenging remains for both the description of 
threshold states.

\section*{Acknowledgements}
I acknowledge the financial support obtained inside the Italian
MIUR program  ``incentivazione alla mobilit\`a di studiosi stranieri e
italiani residenti all'estero''.

\end{document}